\documentclass[twocolumn,twoside,slac_two]{revtex4}
\usepackage{graphicx}
\usepackage{fancyhdr}
\def\mtt{ }
\def\mtta{}
\def\ggg{$\gamma$-ray }
\def\es{}

\def\esss{}
\def\eess{  }
\def\eees{}
\def\mtt{  }
\def\corr{}

\begin{document}

\title{The Surprising Crab Nebula}

%

\author{E. Striani, M. Tavani, V. Vittorini}
\affiliation{INAF/IASF-Roma, I-00133 Roma, Italy}

\begin{abstract}
We will present our study of the {\mtta flux and spectral} variability of the Crab
above 100 MeV on different timescales ranging from days to weeks.
{\mtt In addition to} the four main {\mtt intense and day-long}
flares detected by AGILE and Fermi-LAT between Sept. 2007 and
Sept. 2012, we {\mtt find evidence {\corr for week-long} and less
intense episodes of} enhanced gamma-ray emission that we call
``waves". {\mtt Statistically significant  ``waves'' show
timescales of 1-2 weeks, and can occur by themselves or in
association with shorter flares.} The Sept. - Oct. 2007 gamma-ray {\mtt
enhancement episode detected} by AGILE shows {\mtt both
``wave'' and flaring behavior.}  {\mtt We extend our analysis to
the publicly available Fermi-LAT dataset and show that several
additional ``wave'' episodes can be identified. }
{\mtt We discuss the spectral properties of the September 2007
``wave''/flare event and show that the physical properties of the
``waves'' are intermediate between steady and flaring states.
Plasma instabilities inducing ``waves'' appear to involve spatial
distances $ l \sim 10^{16} \,$cm and enhanced magnetic fields
{\esss $B \sim (0.5 - 1)\,$}mG. Day-long flares are characterized
by smaller distances and larger local magnetic fields. {\mtta
{\corr Typically, the deduced total energy associated with the
``wave'' phenomenon  ($E_w  \sim 10^{42} \, \rm erg$, where $E_w$
is the kinetic energy of the emitting particles) is comparable with}
that associated to the flares, and can reach a few percent of the
total available pulsar spindown energy.  {\corr Most likely,
flares and waves are the product of the same class of plasma
instabilities that we show acting on different timescales and
radiation intensities.} }}

\end{abstract}

\maketitle

\thispagestyle{fancy}

\section{Introduction}

The Crab Nebula (the remnant of a Supernova explosion witnessed by
Chinese astronomers in 1054)  {\mtt is powered by} a very powerful
pulsar (of period $P=0.33$ ms, and  spindown luminosity $L_{sd}
\simeq 5 \times 10^{38} \, \rm erg \, s^{-1} $) (see e.g., Hester
[2008]). {\mtt The pulsar} is energizing the whole system through
the interaction of the particle and wave output within the
surrounding Nebula {\corr(of average magnetic field $\sim 200 \mu$G)}.
The {\mtt resulting unpulsed emission} from
radio to gamma rays up to $100$ MeV is interpreted as synchrotron
radiation from {\mtt at least} two populations of
electrons/positrons energized by the pulsar wind and by
surrounding shocks or plasma instabilities (e.g.,
Atoyan \& Aharonian [1996], Meyer et al. [2010]). The optical and
X-ray brightness enhancements observed in the inner Nebula, known
as ``wisps'', ``knots'', and the ``anvil'' aligned with the pulsar
``jet'' (Scargle [1969]; Hester [1995], [2008]; Weisskopf [2000]),
show {\mtt flux} variations on timescales of weeks or months.
{\mtt On the other hand,} the {\mtt average} unpulsed emission
from the Crab Nebula was always considered essentially stable.
{\mtt  The surprising discovery by the AGILE satellite of variable
gamma-ray emission from the Crab Nebula in Sept. 2010
\citep{tavani1, tavani3}, and the Fermi-LAT confirmation
\citep{buehler, abdo2}  started a new era of investigation of the
Crab system. As of Sept. 2012 we know of four major gamma-ray
flares from the Crab Nebula detected} by
{\corr the AGILE Gamma-Ray Imaging Detector (GRID)} and Fermi-LAT:
\textit{(1)} the Sept-Oct. 2007 event,
\textit{(2)} the Feb. 2009 event,
\textit{(3)} the Sept. 2010, and \textit{(4)}  the
``super-flare'' event of Apr. 2011
(\cite{atel-1}; \cite{atel-2}; \cite{atel-3}; \cite{atel-4}; \cite{striani11}; \cite{2012ApJ...749...26B}).

  {\mtta In this proceeding we address the issue of the gamma-ray
  variability of the Crab Nebula on different timescales ranging
  from days to weeks. We then enlarge the parameter space sampled
  by previous investigations especially for the search of
  statistically significant enhanced emission on timescales of 1-2
  weeks. We will present a brief overview of the current
  knowledge on Crab's main gamma-ray flares and the results of
  a search of \ggg enhanced emission on timescales of weeks in the
  AGILE and Fermi-LAT database and a brief discussion of our results.
  This proceeding is based on \cite{2013ApJ...765...52S},
  where the data analysis and discussion are presented with more details.}

\begin{table*}
  \begin{center}
    \caption{Table of the \emph{flares} (F $\geq 700 \times 10^{-8} \rm \, ph \, cm^{-2} \, s^{-1}$)
  of the Crab Nebula found in the AGILE and Fermi data from Sept. 2007.}\label{tab1}
  \begin{tabular}{|c|c|c|c|c|c|c|c|c|}
    \hline
    & Name &  MJD & \textbf{$\tau_{1}$} (hr) & \textbf{$\tau_{2}$} (hr) & Peak Flux & $B (mG)$ & $\gamma^{\ast}$ ($10^{9}$) &  $l$ ($10^{15}$ cm)\\
    \hline
  2007  & $F_{1}$   & 54381.5   & {\eess $22\pm 11$}    & {\eess $10\pm 5$}  & $1000\pm 150 $     & $ 1.0 \-- 2.0 $     & $2.6\--4.8$    &$1.2\--3.6$\\
   (AGILE)  & $F_{2}$   & 54382.5   & {\eess $14\pm 7$}     & {\eess $6\pm 3$}   & $1400 \pm 200 $    &  {$\esss 1.1\-- 2.1$}    & ${\esss 2.3\--4.3}$   & $0.8\--2.2$\\
            & $F_{3}$   &54383.7    & {\eess $11\pm 5$}    & {\eess $14\pm7$}    & $900 \pm 150$            & $1.0 \-- 2.0$    & $2.6\--4.8$  &$0.8\--1.7$\\ \hline
  2009      & $F_{4}$   &54865.8    & {\eess $10\pm 5$}      & {\eess $20\pm 10$}   &  $700 \pm 140 $   & {$\esss 0.7\--1.3$}    & $2.6\--4.8$    &$0.6\--1.6$ \\
  (FERMI)   & $F_{5}$   &54869.2    & {\eess $10\pm 5$}   & {\eess $22\pm 11$}    & $830 \pm 90 $  & {$\esss 0.8\-- 1.4$} & $2.6\--4.8$      &$0.6\--1.6$\\
   \hline
   2010     & $F_{6}$   &55457.8    & {\eess $8\pm 4$}    & {\eess $22\pm 11$}      & $850 \pm 130 $    & $0.7\-- 1.3$    & $2.5\--4.7$    &$0.5\--1.3$\\
   (AGILE \& & $F_{7}$   &55459.8 & {\eess $6\pm 3$}    & {\eess $6\pm 3$}     & $1000 \pm 100 $   & $ 1.4\--2.6 $   & $2.6\--4.8$     &$0.3\--0.9$\\
      FERMI)      & $F_{8}$   &55461.9  & {\eess $19\pm 10$}  & {\eess $8\pm 4$}   & $750 \pm 110 $   &  {$\esss 0.8\--1.4$}    & $2.5\--4.8$   & $0.9\--3.1$ \\ \hline
  2011      & $F_{9}$   &55665.0    & {\eess $9\pm 5$}     & {\eess $9\pm 5$}        & $1480 \pm 80 $    & $ 1.2\--2.2 $   & $2.8\--5.0$    & $0.5\--1.5$ \\
  (FERMI \&  & $F_{10}$  &55667.3  & {\eess $10\pm 5$}  & {\eess $24\pm 12$}  & $2200 \pm 85 $  & {$\esss 1.3\--2.3$}  & ${\esss 2.7\--4.9}$  & $0.6\--1.6$\\
  AGILE) & & & & & & & &\\ \hline
    \hline
  \end{tabular}
    \end{center}
\noindent {\mtt The timescales $\tau_1$ and $\tau_2$ are the rise
and decay timescales of the flares modelled with {\eess an
exponential fit}
respectively. The {\esss
characteristic length} of the emitting region is {\eees deduced}
{\corr from the relation $l=c \, \delta \, \tau_1$}. The Lorentz factor
$\gamma^{\ast}$ characterizes the adopted model of the  accelerated
particle distribution function $d \, n/d \, \gamma = {\esss
K/\alpha} \cdot \delta(\gamma - \gamma^{\ast})$, {\esss where $K$ is
defined in the spherical approximation. $\alpha=1$ in the
spherical case, and $\alpha < 1$ for cylindrical or pancake-like
volumes reproducing the current sheet geometry.}
{\corr The peak photon flux above 100 MeV is measured in units of
$ 10^{-8} \rm \, ph \, cm^{-2} \, s^{-1}$}.
{\corr These parameters are obtained,
by means of a multi-parameter fit, from the following quantities (in the observer frame):
the position of the peak photon energy, $E_p \propto \delta \gamma^{\ast 2} B$,
the peak emitted power $\nu F \propto \delta^4 K/\alpha \, l^3 B^2 \gamma^{\ast 2}$,
the rise time $\tau_1 = l/(c \delta)$, and the cooling time
$\tau_2 = 8.9 \times 10^3/[(B/\rm Gauss)^2 \, \gamma^{\ast 2} \delta]$, assuming $\delta=1$ (see text).}}
\end{table*}

\begin{figure*}
\begin{center}
\vspace*{-0.3cm} \hspace*{-0.7cm}
 \includegraphics[width=16cm]{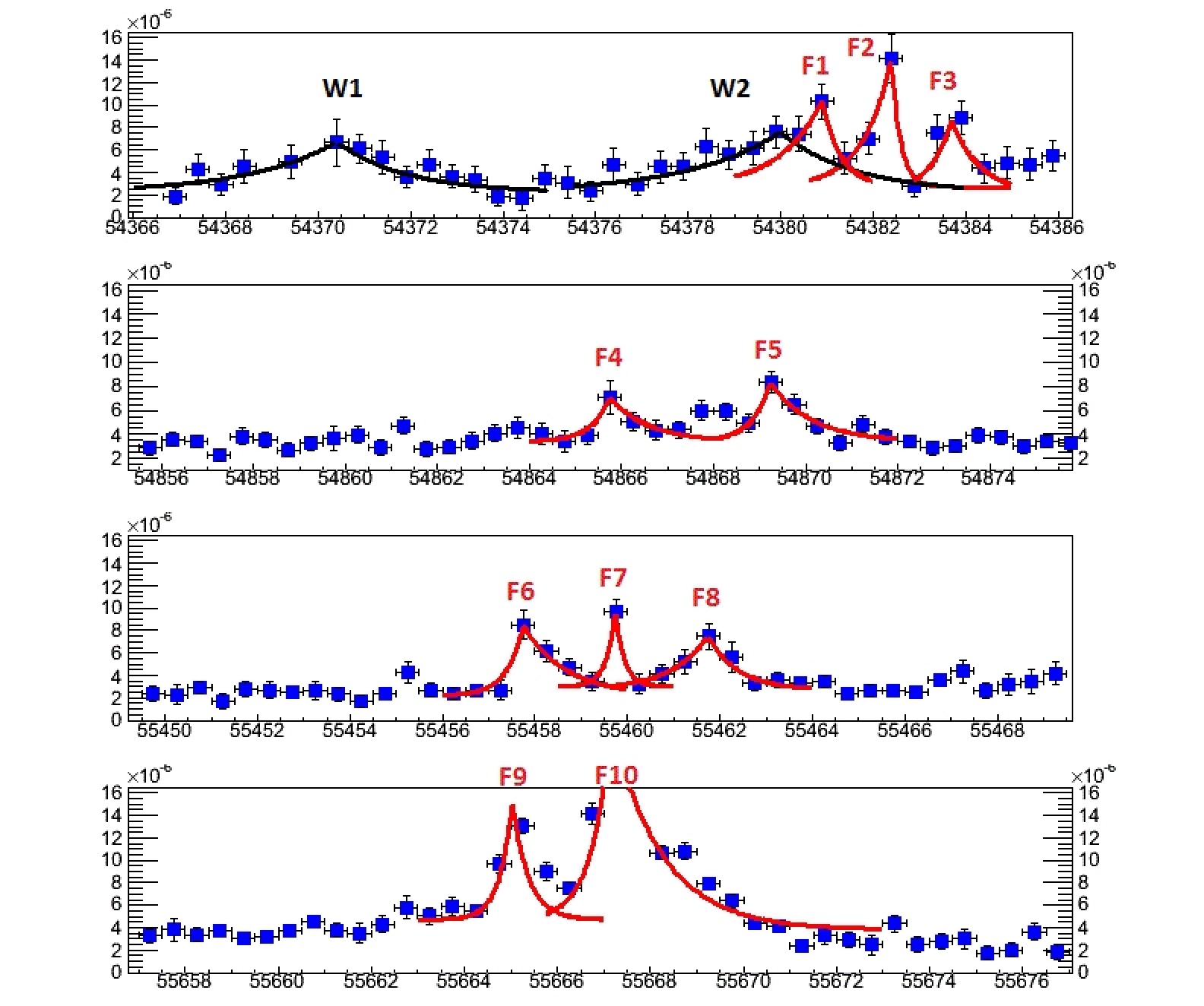}
\caption{Gamma-ray lightcurves above $100$ MeV (12-hr time bins)
from the Crab (pulsar plus Nebula) detected by AGILE and
Fermi-LAT. From top to bottom, the Sept. - Oct. 2007 event (AGILE
data), the Feb. 2009 event (Fermi-LAT data), the Sept. 2010 event
(Fermi-LAT data
), and the Apr. 2011
event (Fermi-LAT data
).}
\label{all_events}
\end{center}
\end{figure*}

\begin{figure*}[t!]
\begin{center}
\vspace*{-0.3cm} \hspace*{-0.7cm}
 \includegraphics[width=16cm]{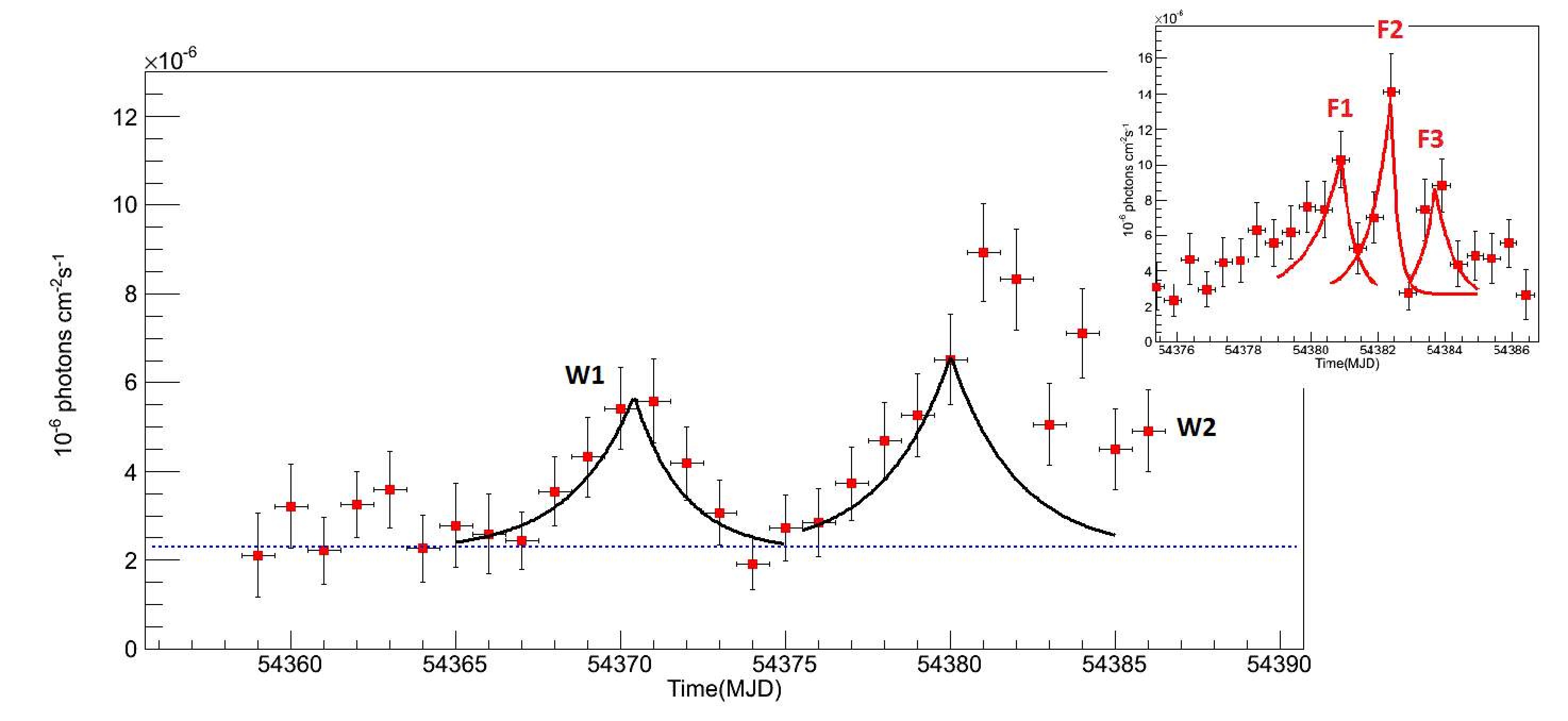}
\caption{Lightcurve (1-day bin) of the Sept.- Oct. 2007 Crab Nebula
flare detected by AGILE. {\es In the inset the 12-hr bin
lightcurve around the flare.} This episode is characterized by a
very strong variability, with waves (black line, marked with a
\emph{W}) and flares (red line, marked with an \emph{F}.)}
\label{flare_2007}
 \end{center}
 \end{figure*}

\section{Overview of the main gamma-ray flares}

{\mtt Four major episodes of intense gamma-ray flaring from the
Crab Nebula have been detected by AGILE and Fermi-LAT
\citep{tavani3,abdo2,striani11,vittorini,2012ApJ...749...26B}.
The definition of a ``flare'' adopted in this paper is that of a
single gamma-ray enhancement event with a risetime $\tau_1 < 1$
day and flux $ F > 700 \times 10^{-8} \rm \, ph \, cm^{-2} \, s^{-1}$ above 100 MeV.
Table 1 summarizes the flaring  events that we find by considering
the AGILE and the publicly available Fermi-LAT database.
{\mtta These events show a complex time structure, being composed
of several sub-flares} that we classify as $Fn'$, with $n'$ a
progressive number. Fig. 1 summarizes the four major flaring
episodes  with the same temporal and flux scales. The colored
curves are indicative of the flaring behavior that in most cases
can be represented by {\eess an exponential fit, characterized by rising ($\tau_1$)
and decay ($ \tau_2$) {\mtta timescales} as given in Table 1. We
also report in Table 1 the data fitting physical parameters
(average local magnetic field $B$, typical particle Lorentz
factor $\gamma^{\ast}$, {\corr and characteristic size $l$ of the emitting region}
of the flaring episodes). {\corr We determine these parameters
and their uncertainties from the time constants $\tau_1$ and $\tau_2$}.
Whenever applicable (2010 and 2011 events),
the AGILE and Fermi-LAT data are consistent both in flux and spectral properties. }

\section{The September-October 2007 event detected by AGILE}

Fig.~\ref{flare_2007} shows the 1-day lightcurve ($E > 100 $MeV) of
the Crab (pulsar plus Nebula) between Sept. 24, 2007 and Oct. 13, 2007.
{\es with a zoom of the lightcurve focused on the short variability episodes in the inset.}
The $\chi^2$ analysis of this data show that there are 2 regions of enhanced emission
that are more than 5 sigma (in a 7-day timescale) above the Crab average flux
in the AGILE data.
These regions have an average flux of $\sim {\eees450} \times 10^{-8} \rm \, ph \, cm^{-2} \,
s^{-1}$ and a rise and decay time of the order of {\mtt several}
days.
{\mtta These slow  components of enhanced \ggg emission show
features different than those of flares that typically have} rise
and decay times of the order of 12-24 hr, and peak fluxes ranging
from $F_{p,5} \simeq 800  \times 10^{-8} \rm \, ph \, cm^{-2} \, s^{-1}$
up to $F_{p,10}\simeq 2500  \times 10^{-8} \rm \, ph \, cm^{-2} \, s^{-1}$
{\mtt (as for {the Crab super-flare of Apr. 2011)}}.
We call these slow component of enhanced emission \emph{waves}.
We indicate with $W_1$ the emission from MJD $\simeq$ 54367 to
MJD $\simeq$ 54374 and with $W_2$ the emission from MJD $\simeq$ 54376 to MJD $\simeq$ 54382.
{\es The 12-hr lightcurve in the inset of Fig.~\ref{flare_2007} shows that
the 2007 {\mtt peak intensity
event, that in our previous 
analysis \citep{tavani3} appeared unresolved}, 
is actually composed of three different {\mtt flaring components},
that we indicate by $F_1$, $F_2$, and $F_3$}.
{\corr Peak fluxes, rise times and decay times (estimated with an exponential fit)
 for $F_1$, $F_2$, and $F_3$ and for $W_1$ and $W_2$ are presented in Tab.~\ref{tab1}.
The post-trial probability, $P_{post}=1-(1-p)^{N_t}$,
{\mtt with} $p$ the pre-trial
probability {\corr obtained from the $\chi^2$ test},
and $N_t$ the number of 7-day maps (trials), is $P_{post}> 5\sigma$ for
both $W_1$ and $W_2$.}
{\corr We find that the differential} particle energy distribution
function (per unit volume) can be described by a monoenergetic function,
 $d \, n / d \, \gamma
= {\esss K/\alpha} \cdot \delta(\gamma-\gamma^{\ast})$, where $\gamma$
is the particle Lorentz factor, {\eees and} $\gamma^{\ast}$ is the monochromatic
value of the particle energy. {\mtta The electron density constant $K$} is defined
in the spherical approximation,
 with $\alpha=1$ in the spherical case,
  and $\alpha < 1$ for cylindrical or pancake-like volumes.
A power-law distribution and/or a relativistic
Maxwellian distribution were {\corr also} shown to be consistent with the
flaring data (Tavani et al., [2011a], Striani et al. [2011b]). We adopt
{\mtta here} a monochromatic distribution that deconvolved with
the synchrotron emissivity leads to an emitted spectrum
practically indistinguishable {from} the relativistic Maxwellian
shape (e.g., Striani et al. [2011b], Buehler et al. [2012]). This
distribution is in agreement with all available gamma-ray data
(for both flaring and ``wave'' behavior, see below), and reflects
an important property of the flaring Crab acceleration process
(Tavani [2013]). }
With our calculations {\corr we find for $W_1$ a magnetic field
$B=(0.8\pm0.2)$ mG, a Lorentz factor $\gamma^{\ast}=(4\pm1) \times 10^9$,
and a typical emitting {\esss
length range $l = (0.5 \-- 1.5) \times 10^{16}$ cm}. For $F_2$,
we find $B = (1.5\pm0.5$) mG, $\gamma^{\ast} = (3\pm1)\times 10^9$,
and $l = (1.5\pm 0.7) \times
10^{15}$ cm.}
{\mtt In our model, the total number of accelerated particles
producing the gamma-ray wave/flaring behavior is in the range
{\esss $N \sim (1 - 3)\cdot 10^{38} (\Delta \Omega/4\pi$)},
{\mtta with $\Delta \Omega$ the solid angle of the \ggg emission}.
It is interesting to note that besides the differing values of the
magnetic field and particle densities, the typical Lorentz factor
{\corr and total number of radiating particles are} similar for the ``wave'' $W_1$ and the flare $F_2$. We
find that this {\eess is} a typical behavior of the transient gamma-ray
emission that appears to be well represented by a {\corr monochromatic
particle distribution function with $\gamma^{\ast} \simeq (3\--5) \times
10^9$. }}

\section{Search for enhanced gamma-ray emission in the Fermi-LAT data}

Motivated by the waves found in the AGILE data, we searched for a
similar type of enhanced gamma-ray emission in the {\mtt publicly
available} Fermi-LAT data.
\begin{figure*}
\begin{center}
\vspace*{-0.3cm} \hspace*{-0.7cm}
 \includegraphics[width=16 cm]{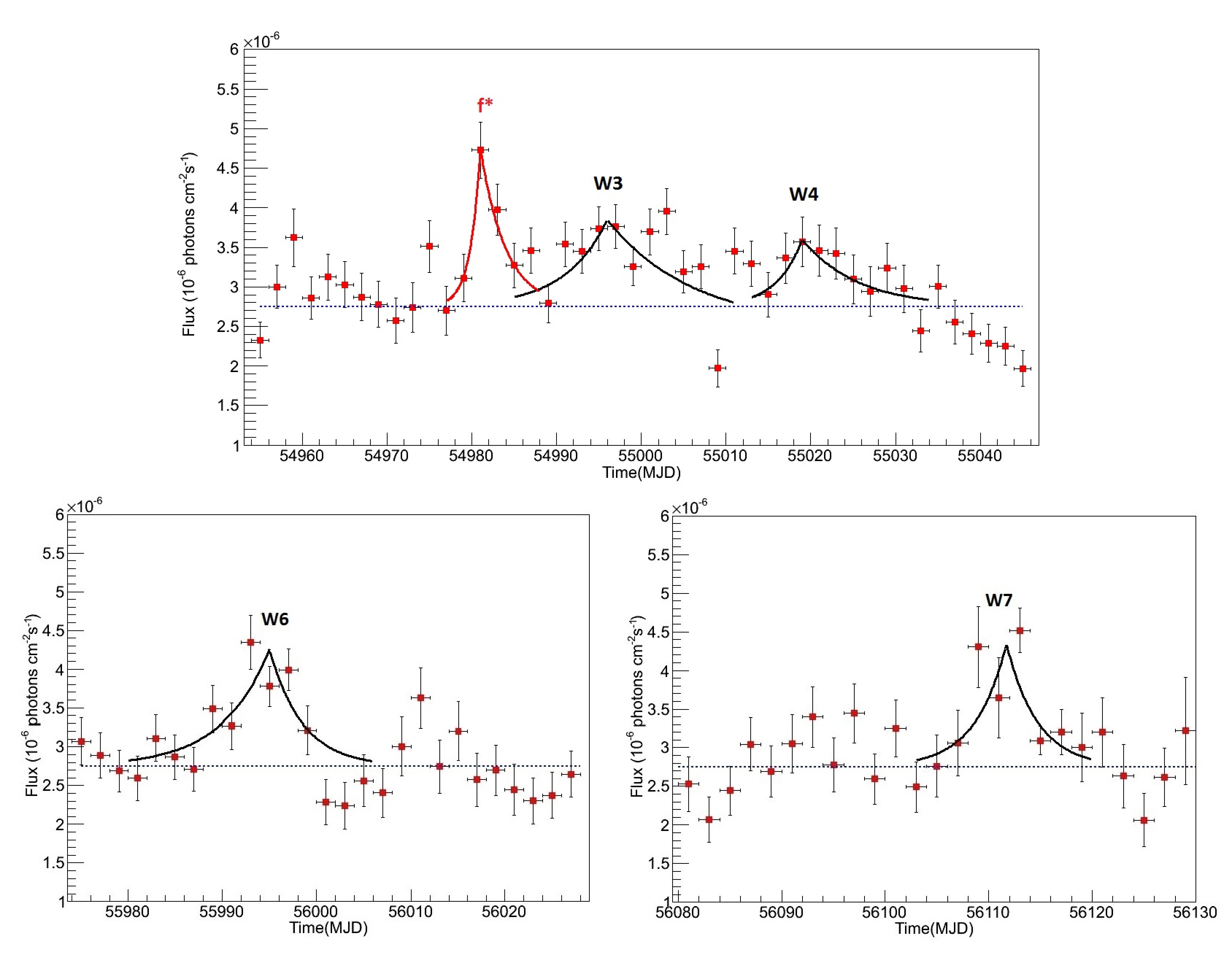}
\caption{ \mtt The most prominent \emph{wave} episodes in the
Fermi-LAT data with more than $5\sigma$ enhancements  above the
Crab average emission. {\eess 2-day binned gamma-ray flux values
(in unit of $10^{-6} \, \rm ph \, cm^{-2} \, s^{-1}$) above 100
MeV as a function of time. We notice that the event marked as $f^{\ast}$ is intermediate between flares and waves.}} \label{Fermi_5sigma}
 \end{center}
 \end{figure*}
With our $\chi^2$ analysis of the Fermi data we identified
four episodes of enhanced gamma-ray emission that are, in a 8-days timescale,
more than 5-sigma above the Crab average flux.
We call them $W_3, W_4, W_6$ and $W_7$
 in Fig.~\ref{Fermi_5sigma}. These episodes have apparent
durations in the range from 8 to 50 days (see the complex marked
as $W_3-W_4$), and an average flux in the range $F =
(350-500)\times 10^{-8} \rm \, ph \, cm^{-2} \, s^{-1}$.
The {\mtta event  $W_7$ is coincident with the {\mtt enhancement episode}  detected\footnote{Due to solar panel
constraints, this event was unobservable by AGILE.} by Fermi-LAT
(Ojha et al., 2012).}
{\mtt Fig.~\ref{Fermi_5sigma} shows the detailed lightcurves of
the Fermi-LAT wave episodes with the largest post-trial
significance}.
{\mtt Fig.~\ref{Fermi_5sigma}  shows a remarkable episode of
{\mtta ``wave''} enhanced emission near MJD = 55000}. The Crab was
for about 50 days above $5\sigma$ from its standard gamma-ray
flux, with an average flux in this period $F = (340 \pm 6)
\times 10^{-8} \rm \, ph \, cm^{-2} \, s^{-1}$. {\mtta This event
{\mtta appears} quite complex\footnote{{\mtt For simplicity, we
use {\eess an exponential} approximation for the ``wave'' emission. Admittedly,
{\eess this} approximation for the episode of Fig.~\ref{Fermi_5sigma}
centered on MJD $= 55000$ is not adequate. It is shown in
Fig.~\ref{Fermi_5sigma} for illustrative purposes only.}}.
The total {\mtt episode} that includes
$f^{\ast} $, $W_3$ and $W_4$ has a time duration of $\sim 50$
days, a pre-trial probability $p=2\times 10^{-25}$ and a
post-trial significance $\sigma_{post}>10$.
{\es For each ``wave'' that we found at $5\sigma$ above the Crab
average emission, we estimated the rise $\tau_1$ and the decay
time $\tau_2$, the average flux, the peak flux, the probability of
obtaining the given $\chi^2$ in the null hypothesis, and the post-trial
significance. The rise and decay timescales are estimated
{\eess with an exponential fit.}
The results are summarized in Tab. 2.

\begin{table*} 
\begin{center}
  \small
   \caption{Table of the \emph{waves} above $5\sigma$ post-trial from the Crab average emission found in the AGILE and Fermi data.}
  \begin{tabular}{|c|c|c|c|c|c|c|c|c|}
  \hline
    Name & MJD & Duration & $\tau_1$ & $\tau_2$ & Average Flux & Peak Flux &  Pre-trial & Post-trial \\
   &  & (days) & (days) & (days) & $ (10^{-8} \rm \, ph \, cm^{-2} \, s^{-1})$  &  $ (10^{-8} \rm \, ph \, cm^{-2} \, s^{-1})$   & p-value & significance\\ \hline
  $W_{1}$ & 54368-54373     & 5      &  {\eess $2\pm 1$}   & {\eess $2\pm 1 $}     &$440\pm40$  &$670\pm200$   &$4.5 \times 10^{-8}$   & 5.0 \\\hline
  $W_{2}$ & 54376.5-54382.5 & 6      & {\eess $2.5\pm 1$}  & {\eess $2\pm 1 $}     &$480\pm40$  &$760\pm140$   &$3.0 \times 10^{-9}$   & 5.5\\ \hline
  $W_{3}$ & 54990-55008     & 18     & {\eess $5\pm 2.5$}    & {\eess $10\pm 5$}    &$352\pm9$   &$380\pm30$    &$1.0 \times 10^{-8}$    & 4.6\\ \hline
  $W_{6}$ & 55988-56000      & 12     & {\eess $5\pm 2.5$}    & {\eess $3.5\pm 1.5$}   &$367\pm12$  &$435\pm35$    &$1.8 \times 10^{-12}$    & 6.2\\ \hline
  $W_{7}$ & 56108-56114      & 6      & {\eess $3\pm 1.5$}    & {\eess $3\pm 1.5$}     &$431\pm22$  &$450\pm30$    &$1.9 \times 10^{-9}$   & 5.9\\ \hline
  \hline
\end{tabular}
\end{center}
\noindent {Photon fluxes are obtained for $E_{\gamma}> 100$ MeV.}
  \label{tab2}
\end{table*}

\section{Constraints from a  synchrotron cooling model}

{\mtta We can deduce important physical parameters of the enhanced
\ggg emission by adopting a synchrotron cooling model (see also
Tavani et al. [2011a], Vittorini et al. [2011], Striani et al. [2011b]).
Tables 1 and 2 summarize the
relevant information for the major ``flares'' and ``waves''.}
{\mtta We find that for flares lasting 1-2 days the typical length
is  $l \simeq (1 - 2)\times 10^{15}$ cm, the density constant in the
range $K/\alpha = (2 - 8)\times 10^{-9} \, \rm cm^{-3}$, the
typical Lorentz factor $\gamma^{\ast} =(2.5\--4.5) \cdot 10^{9}$, and the
local magnetic field affecting the cooling phase in the range $B =
(1-2) $ mG. As discussed above, the total number of radiating particles,
{\corr in case of unbeamed ($\delta=1$) isotropic emission,
is \corr $N\sim (1\--3) \cdot 10^{38}$} .}

{\mtta For the ``wave'' episodes, the typical length is  $l >
10^{16}$ cm, the density constant is in the range $K/\alpha = (2 -
8)\times 10^{-11} \, \rm cm^{-3}$, {\corr the  Lorentz factor $\gamma^{\ast}
=(3-5) \cdot 10^{9}$}, and the local magnetic field in the range
$B = (0.5-1)$ mG. The total number of particles involved is of
the same order as in the case of flares.

\section{Discussion and Conclusions}

By considering AGILE and Fermi-LAT gamma-ray data above 100 MeV we
find that the Crab produces a broad variety of enhanced emission.
We characterize this enhanced emission as short timescale (1-2
day) ``flares'' and  long timescale (1 week or more) ``waves''.
Given the current detection level of 1-2 day enhancements (which
can be extended to longer timescales of order of 1-2 weeks), we
cannot exclude that the Crab is producing an even broader variety
(in flux and timescales) of enhanced gamma-ray emission. With the
current sensitivities of {\corr $\gamma$-ray telescopes} we can
explore an important but necessarily limited range of flux and
spectral variations.
{\corr It is interesting to note that
what we called ``flares'' and ``waves'' (a somewhat arbitrary
division) share the same spectral properties. Given the current
gamma-ray sensitivities, we could have detected different spectral
behaviors in the energy range 50 MeV - 10 GeV. Most likely, flares
and waves are the product of the same class of plasma
instabilities that we show acting on different timescales and
radiation intensities.
Whether or not the instability driver of this process is truly
stochastic in flux and timescales will be determined by a longer
monitoring of the Crab Nebula.}
The pulsar wind outflow and nebular interaction conditions need to
be strongly modified by instabilities in the relativistic flow
and/or in the radiative properties. Plasma instabilities possibly
related to magnetic field reconnection in specific sites in the
Nebula can be envisioned. However, {\mtta evidence for magnetic
field reconnection events in the Crab Nebula is elusive, and } no
 optical or X-ray emission in coincidence with the gamma-ray
flaring has been {\mtta unambiguously} detected to date (e.g.,
Weisskopf et al. [2012]).

Both the flaring and ``wave'' events
can be attributed to a population of accelerated electrons consistent with a
mono-chromatic or relativistic Maxwellian distribution of
typical energy $\gamma^{\ast} \sim (2.5\--5) \cdot 10^9$.

We also notice that the emitted total energies that can be deduced
for the wave ($\delta E_{\gamma,w}$) and flare ($ \delta
E_{\gamma,f}$) episodes in general satisfy the relation $\delta
E_{\gamma,w} \simeq \delta E_{\gamma,f}$. The total gamma-ray emitted energy for the wave episode
$W_1$ can be {\corr estimated as $\delta E_{\gamma,w1} \sim 10^{41} \, \rm erg$}.
Therefore, the {\corr energy associated with the } wave
event $W_1$, taken here as an example of Crab ``wave'' emission,
{\corr can reach a few percent} of the total spindown energy.

We conclude that the Crab ``wave'' events are highly significant
and quite important from the energetic point of view. ``Waves''
typically imply regions larger than in the case of flares, and
smaller average magnetic fields. Their total emitted
gamma-ray energy can be {\corr comparable with} that associated with shorter
flares. More observations of this fascinating phenomenon are
necessary to improve our knowledge of the flaring Crab.

\bigskip 

\begin{thebibliography}{9}   


\bibitem[Abdo et al.(2011)]{abdo2} Abdo, A.A., et al., 2011, Science, 331, 739.

\bibitem[Atoyan \& Aharonian(1996)]{atoyan} Atoyan, A.M. \& Aharonian, F.A., 1996, MNRAS, \textbf{278}, 525.

\bibitem[Buehler et al. (2010)]{buehler} Buehler, R., \textit{et al.}, 2010, Astron. Telegram 2861.

\bibitem[Buehler et al. (2011)]{atel-1} Buehler, R., \textit{et al.}, 2011, Astron. Telegram 3276.

\bibitem[Buehler et al. (2012)]{2012ApJ...749...26B} Buehler, R., Scargle, J.~D., Blandford, R.~D., et al.\ 2012, ApJ, 749, 26

\bibitem[Hays et al. (2011)]{atel-3} Hays, E., \textit{et al.}, 2011, Astron. Telegram 3284.

\bibitem[Hester, Scowen \& Sankrit (1995)]{hester2} Hester, J.J., P. A. Scowen \& R. Sankrit \textit{et al.}, 1995,
ApJ, \textbf{448}, 240.

\bibitem[Hester, Mori \& Burrows, (2002)]{hester3} Hester, J.J.,
Mori, K., Burrows, D. \textit{et al.}, 2002, ApJ, \textbf{577},
L49.

\bibitem[Hester (2008)]{hester1} Hester, J.J., 2008, Annual Rev. Astron. \& Astrophys., \textbf{46}, 127.


\bibitem[Meyer, Horns \& Zechlin(2010)]{meyer} Meyer, M., Horns, D.
\& Zechlin, H.S., 2010, A\&A, 523, A2.

\bibitem[Ojha et al.(2012)]{ojha} Ojha, R., et al., 2012, Astron. Telegram 4239.

\bibitem[Pittori et  al.(2009)]{2009A&A...506.1563P} Pittori, C., Verrecchia, F., Chen, A.~W., et al.\ 2009, A\&A, 506, 1563


\bibitem[Scargle (1969)]{scargle} Scargle, J.D., 1969, ApJ, \textbf{156}, 401.

\bibitem[Striani et al.(2013)]{2013ApJ...765...52S} Striani, E., Tavani,
M., Vittorini, V., et al.\ 2013, ApJ, 765, 52

\bibitem[Striani et al. (2011a)]{atel-4} Striani, E., \textit{et al.}, 2011a, Astron. Telegram 3286.

\bibitem[Striani et al. (2011b)]{striani11} Striani, E., Tavani, M., Piano, G., et al.\ 2011b, ApJL, 741, L5

\bibitem[Tavani et al. (2010)]{tavani1} Tavani, M. \textit{et al.}, 2010, Astron. Telegram 2855.

\bibitem[Tavani et al. (2011a)]{tavani3} Tavani, M., \textit{et al.}, 2011a, Science, \textbf{331}, 73.

\bibitem[Tavani et al. (2011b)]{atel-2} Tavani, M., \textit{et al.}, 2011b, Astron. Telegram 3282.

\bibitem[Tavani(2013)]{tavani4} Tavani, M., 2013, in preparation.

\bibitem[Vittorini et al. (2011)]{vittorini} Vittorini, V., \textit{et al.}, 2011, ApJ, \textbf{732}, L22.

\bibitem[Weisskopf et al. (2000)]{weisskopf} Weisskopf, M.C., Hester, J.J., A. F. Tennant, R. F. Elsnor, N. S. Schulz \textit{et al.}, 2000,
ApJ, \textbf{536}, L81.

\bibitem[Weisskopf et al.(2013)]{2013ApJ...765...56W} Weisskopf, M.~C.,
Tennant, A.~F., Arons, J., et al.\ 2013, ApJ, 765, 56




\end{thebibliography}

\end{document}